%% file: main.tex
\documentclass[acmsmall]{acmart}

\usepackage{amsmath, amsfonts, amsthm, enumerate}
\usepackage{graphicx}
\usepackage{textcomp}
\usepackage{mathrsfs}
\usepackage{multirow}
\usepackage{array}
\usepackage{subfigure}
\usepackage{subcaption}
\usepackage{footnote}
\usepackage{paralist}
\usepackage{float}
\usepackage{threeparttable}
\usepackage{booktabs}
\usepackage{rotating} 
\usepackage{color,framed}
\usepackage{setspace}
\usepackage{balance}
\usepackage{tcolorbox}
\usepackage[justification=centering]{caption}
\usepackage{url}
\usepackage{epstopdf}
\usepackage{listings}

\newcommand{\CodeIn}[1]{{\small\texttt{#1}}}

\newcommand{\benchmark}{\emph{InterCode}}
\newcommand{\stepone}{Prgramming-Task Collection}

\usepackage{xspace}
\usepackage{balance}
\usepackage{pifont}
\usepackage{multirow}
\usepackage{hyperref}
\usepackage{breakurl}
\usepackage{cleveref}
\usepackage{balance}
\usepackage{amsmath}
\usepackage{indentfirst}
\usepackage{listings}
\usepackage{tabularx}
\usepackage{subfigure}
\usepackage{booktabs}
\usepackage{xcolor}
\usepackage{multirow}
\usepackage{tcolorbox}
\usepackage{bm}
\usepackage{microtype}
\usepackage[linesnumbered,ruled,vlined,noend]{algorithm2e}
\usepackage{algpseudocode}

\usepackage{fancyhdr}

\newcommand{\mr}[1]{\textcolor{black}{#1}}

\begin{document}

\title[An Evaluation of Large Language Models on Iterative Example-Based Code Generation]{The First Prompt Counts the Most! An Evaluation of Large Language Models on Iterative Example-Based Code Generation}

\author{Yingjie Fu}
\orcid{0000-0003-2574-9774}
\affiliation{%
  \institution{School of Computer Science, Peking University}
  \city{Beijing}
  \country{China}
}
\email{yingjiefu@stu.pku.edu.cn}

\author{Bozhou Li}
\orcid{0009-0001-7519-5733}
\affiliation{%
  \institution{Peking University}
  \city{Beijing}
  \country{China}
  }
\email{libozhou@pku.edu.cn}

\author{Linyi Li}
\authornote{Tao Xie~(taoxie@pku.edu.cn) and Linyi Li~(linyi\_li@sfu.ca) are correspondence authors.}
\orcid{0000-0002-5403-3217}
\email{linyi_li@sfu.ca}
\affiliation{%
  \institution{School of Computing Science, Simon Fraser University}
  \city{Burnaby}
  \state{BC}
  \country{Canada}
}

\author{Wentao Zhang}
\orcid{0000-0002-7532-5550}
\email{wentao.zhang@pku.edu.cn}
\affiliation{%
 \institution{Center for Machine Learning Research, Peking University}
  \city{Beijing}
  \country{China}
}

\author{Tao Xie}
\authornotemark[1]
\orcid{0000-0002-6731-216X}
\email{taoxie@pku.edu.cn}
\affiliation{%
  \institution{Key Laboratory of High Confidence Software Technologies (Peking University), Ministry of Education; School of Computer Science, Peking University}
 \city{Beijing}
  \country{China}}

\begin{abstract}
The capabilities of Large Language Models (LLMs) in code generation have been extensively studied, particularly for implementing target functionalities from natural-language descriptions. 
As an alternative to natural language, input-output  (I/O) examples provide an accessible, unambiguous, and flexible way to describe functionalities. However, their inherent diversity, opaqueness, and incompleteness impose greater challenges for understanding and implementing the target requirements. 
Therefore, generating code from I/O examples (i.e., example-based code generation) provides a new perspective, allowing us to additionally evaluate LLMs' capability to infer target functionalities from limited information and to process new-form requirements. However, related research about LLMs in example-based code generation remains largely unexplored.
To fill this gap, this paper presents the first comprehensive study on example-based code generation using LLMs. 
To address the incorrectness caused by the incompleteness of I/O examples, we adopt an iterative evaluation framework and formalize the objective of example-based code generation as two sequential sub-objectives: generating code conforming to the given examples and generating code that successfully implements the target functionalities from (iteratively) given examples.
We assess six state-of-the-art LLMs using a new benchmark of 172 diverse target functionalities (derived from HumanEval and CodeHunt). 
The results demonstrate that when requirements are described using iterative I/O examples rather than natural language, the LLMs' score decreases by over 60\%, indicating that example-based code generation remains challenging for the evaluated LLMs.
Notably, the vast majority (even over 95\%) of successfully implemented functionalities are achieved in the first round of the iterations, suggesting that the LLMs struggle to effectively utilize the iteratively supplemented requirements.
Furthermore, we find that combining I/O examples with even imprecise and fragmental natural language descriptions greatly improves LLM performance, and the selection of initial I/O examples can also influence the score, suggesting opportunities for prompt optimization.
These findings highlight the importance of early prompts during interactions and offer critical insights and implications for enhancing LLM-based code generation. 
\end{abstract}



\maketitle
\input{Introduction.tex}

\input{Background.tex}
\input{Framework.tex}

\input{NewBenchmark.tex}

\input{Evaluation.tex}

\input{Results.tex}

\input{Recommendation.tex}
\input{Related.tex}
\input{Threats.tex}
\input{Conclusion.tex}

\section{Acknowledgment}
This work was partially supported by National Natural Science Foundation of China under Grant No. 92464301. 

\bibliographystyle{ACM-Reference-Format}
\bibliography{reference}
\end{document}

%% file: Introduction.tex
\section{Introduction}
\label{sec:Introduction}
Code generation has been recognized as one of the most important and promising applications of large language models (LLMs)~\cite{LEVER-ICML23}. State-of-the-art LLMs, e.g., Llama~\cite{llama2, llama3, CodeLlama}, Gemma~\cite{gemmateam2024gemma, gemma2, codegemma}, DeepSeek~\cite{DeepSeek-Coder, deepseekv2}, ChatGPT~\cite{ChatGPT}, and GPT4~\cite{GPT4}, have shown impressive capabilities in generating executable programs from prompts detailing target functionalities.
Typically, a prompt for a target functionality consists of a natural-language description, and may sometimes include supplementary information such as input-output (I/O) examples and function signatures~\cite{TACO, APPS, MBPP, Codex, DS-1000, AixBench, MultiPL-E, AlphaCode, ClassEval}.

In addition to natural language, I/O examples also provide an easily accessible, unambiguous, and flexible way to describe the target functionalities. 
\textbf{First}, I/O examples offer a straightforward and user-friendly alternative when natural-language descriptions are unavailable.
For non-expert users who struggle to articulate requirements clearly, I/O examples provide a straightforward way to express their intent~\cite{Gulwani2016PBE}.
For reverse engineering tasks~\cite{Reverse-Synthesis} (e.g., binary de-obfuscation~\cite{deobfuscation}) whose goal is to reproduce existing programs or interfaces with unknown functionalities, I/O examples can be iteratively gathered through interactions to reveal these functionalities. 
\textbf{Second}, I/O examples are concrete and precise, being able to reduce misunderstanding in functional descriptions~\cite{balog2017deepcoder, InductiveProgramming}. 
Specifically, I/O examples clearly illustrate the expected outputs for specific inputs, offering clear guidance on program behavior~\cite{TDD-clear}. 
This characteristic allows I/O examples in functional descriptions to serve as tests directly, enabling automated and efficient correctness checking of the generated code~\cite{TDD-astest, TDD-quick}. 
\textbf{Third}, I/O examples can be dynamically updated to clarify, refine, or expand the target functionalities. For instance, both the failing tests and the observed edge cases during development can be used to create new I/O examples, aiding in the adaptive clarification and refinement of functionality descriptions~\cite{TDD-astest, TDD-improve}.

Functional descriptions in the form of I/O examples present three additional challenges for code generation tasks.
\textbf{First}, I/O examples are not frequently included in the training data for code generation~\cite{wen2024grounding}, posing difficulties for models in understanding the requirements conveyed in this form.
Specifically, I/O examples are often limited in quantity, typically appearing in test cases or supplements to the natural-language descriptions. Compared to the potentially extensive input space, these examples can cover only a small fraction.
\textbf{Second}, I/O examples do not explicitly state how to derive the expected outputs, placing a high demand on the inferring and generalizing capability of a code generator. 
Without hints about the structure or logic of the code, LLMs must deduce the underlying transformation from a limited number of I/O examples and apply them across diverse contexts.
\textbf{Third}, a single set of I/O examples usually cannot completely specify target functionalities, requiring a code generator to iteratively receive supplementary prompts and refine generated code. 
In extreme scenarios, the code may simply match each input with a branch to satisfy the given examples but does not achieve the target functionality. Therefore, it is important for a code generator to utilize adaptively supplemented prompts.


To investigate the potential of LLMs on code generation from I/O examples (aka example-based code generation), in this paper, we conduct the first comprehensive study (to the best of our knowledge). 
Considering the inherent incompleteness (i.e., hardly achieving a comprehensive sampling of the input space) of I/O examples in describing the target functionality, we refine the objective of example-based code generation into two sequential sub-objectives, and propose an iterative evaluation framework to provide supplementary I/O examples adaptively.

\begin{itemize}
    \item \textbf{Sub-Objective1 (O1)}: Generating code that conforms to all given I/O examples. This objective concerns the capability to \textbf{understand requirements conveyed through I/O examples}. 
    Specifically, it focuses on only the given I/O examples, disregarding whether the code satisfies all possible I/O examples of the target functionalities.

    \item \textbf{Sub-Objective2 (O2)}: Generating code that successfully implements the target functionality. This objective additionally concerns two capabilities: \textbf{inferring target functionalities from I/O examples} and \textbf{improving generated code through iterative feedback}. 
    The target functionality is defined by \emph{reference code} that is executable but invisible to LLMs. The generated code is expected to be input-output equivalent to the reference code. Otherwise, the framework adaptively supplements new I/O examples to reveal their differences.
    
\end{itemize}

To enable a comprehensive evaluation, we construct a benchmark comprising 172 target functionalities drawn from existing code benchmarks. Each functionality is accompanied by five sets of randomly sampled I/O examples as the starting point of the iteration.
With this new benchmark, we conduct thorough evaluations and analysis on six state-of-the-art large language models (one closed-source and five open-source).

\textbf{Evaluation Results.} 
First, the evaluation results reveal that when programming requirements are provided in the form of only I/O examples (rather than natural languages), the code generation capability of LLMs declines greatly.
Furthermore, the score for finally implementing the target functionality drops even over 60\%.
Among the evaluated models, GPT-4o-mini achieves pass@10 values ranging from 0.30 to 0.32, outperforming all other open-source models with approximately 7B parameters. Meanwhile, DeepseekCoder-6.7b-instruct achieves pass@10 values between 0.22 and 0.24, leading among open-source models with an approximately 80\% improvement over the second place. 
Moreover, we find that providing I/O examples along with relevant natural language information (even if that information is inaccurate and fragmented) can 
substantially improve scores.
Finally, by analyzing the results under different types of functionalities, we conclude that it is easier for the evaluated LLMs to generate code for functionalities related to string manipulations in example-based code generation. In addition to the functionality-related characteristics, LLMs' scores are also influenced by the selection of I/O examples in the prompt.

\textbf{Findings and Implications.} After identifying the limitations of LLMs in example-based code generation, we further analyze the generated code and trends across iterations. First, we observe that the code generated by LLMs may simply employ \CodeIn{if} statements to match the given I/O examples, and this tendency most commonly occurs with Llama-2-7b-chat. This observation illustrates the necessity of iterative evaluation frameworks for example-based code generation. Second, during the iteration process, the very first rounds of interactions play the most critical role in the ultimate success, because the evaluated LLMs are not good at utilizing the iteratively supplemented feedback. This finding underscores the importance of selecting appropriate initial I/O examples for example-based code generation. More importantly, our benchmark covers an under-explored topic (code generation with multi-turn requirements) and suggests that current LLMs may be relatively weak in achieving multi-turn requirements and iteratively given requirements compared to single-turn ones.

In summary, our paper makes the following main contributions.
\begin{itemize}
    \item \textbf{The first comprehensive study of LLMs' capability in example-based code generation.} We regard example-based code generation as a task with multi-turn requirements, formalizing it into two sequential sub-objectives. 
    \item \textbf{An iterative evaluation framework for example-based code generation and a new benchmark applicable to this framework.} 
    Both the framework and the benchmark can be reused and extended for more programming languages and functionalities.
    \item \textbf{Comprehensive evaluation, comparison, and analysis of six state-of-the-art LLMs.} The evaluation compares the scores of different LLMs, summarizes their trends over iterations, analyzes their strengths/weaknesses in different types of functionalities, and provides an initial exploration of factors that may contribute to improvements.
    \item \textbf{Empirical evidence of current LLMs' limitations in example-based code generation, particularly in handling I/O example requirements and refining code through iterative feedback.} The evidence offers an important insight into LLMs' code generation capability and provides valuable suggestions for applying LLMs to code generation with multi-turn requirements.
\end{itemize}

%% file: Background.tex
\section{Background}
\label{section: background}
In this section, we first introduce example-based code generation, an important area in both the research and practice communities. After that, we summarize the progress of LLMs, highlighting their outstanding capabilities in code generation.

\subsection{Example-based Code Generation}
Example-based Code Generation (aka Programming by Examples, PBE)~\cite{Gulwani2016PBE}, referring to automatically synthesizing programs specified by only input-output examples (I/O examples),
has been widely illustrated to be powerful in many real-world applications~\cite{web-scripts, string-processing, data-extraction, data-extraction2, CAD, Visualization, merge-conflict, document-editing}, e.g., web automation~\cite{web-scripts}, string processing~\cite{string-processing}, and data extraction~\cite{data-extraction, data-extraction2}.
Typical approaches for example-based code generation leverage search-based algorithms~\cite{PBE-search}, which are feasible on only carefully designed domain-specific language. However, for general-purpose programming languages, these approaches struggle with complicated syntax and extensive search space.

\begin{figure*}[t]
\centering
\includegraphics[width=0.98\textwidth]{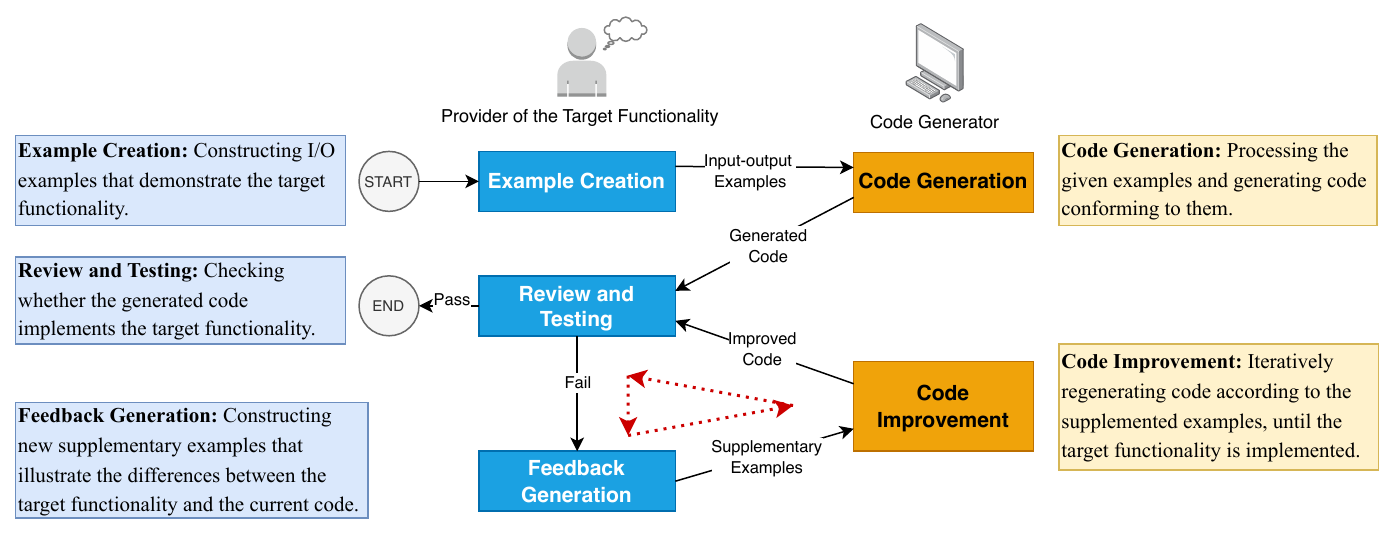}
\caption{The Interactive Workflow of Example-Based Code Generation}
\label{fig: interaction}
\vspace{-2em}
\end{figure*}

I/O examples can serve as an easily accessible and understandable format of specifications, but they are usually incomplete~\cite{incomplete} for describing functionalities. In other words, the generated code may conform to all the given examples but not implement the target functionality. 
As a result, example-based code generation requires an interactive workflow, which allows iterative feedback to clarify the specification. The interactive workflow usually includes five steps (as shown in Figure~\ref{fig: interaction}). 
Inspired by the workflow, our evaluation adopts an iterative evaluation framework, which can also adaptively construct supplementary I/O examples to clarify the target functionality.

\subsection{LLMs on Code Generation}
State-of-the-art LLMs~\cite{ChatGPT, llama2, gemmateam2024gemma} have shown their impressive capabilities in various natural-language tasks~\cite{llm-evaluation-survey}, including code generation (e.g., generating code snippets, completing functions, and even solving competitive programming problems) by understanding natural language prompts of the target functionalities~\cite{DeepSeek-Coder, codegemma, CodeLlama, AlphaCode, Codex, CodeGen}. 
Particularly, even medium-sized LLMs (those with fewer than 10B parameters, e.g., DeepSeek-Coder 6.7b~\cite{DeepSeek-Coder}) can achieve over 50\% correctness on commonly used programming languages.



%% file: Framework.tex
\section{Iterative Evaluation Framework}
\label{sec: interactive framework}
In this section, we introduce the iterative evaluation framework consisting of two stages: first-round interaction and iterative interaction.
Specifically, first-round interaction focuses on generating code that conforms to all the given I/O examples, and iterative interaction focuses on implementing the target functionality according to the (iteratively supplemented) I/O examples.

\begin{figure*}[t]
\centering
\includegraphics[width=\textwidth]{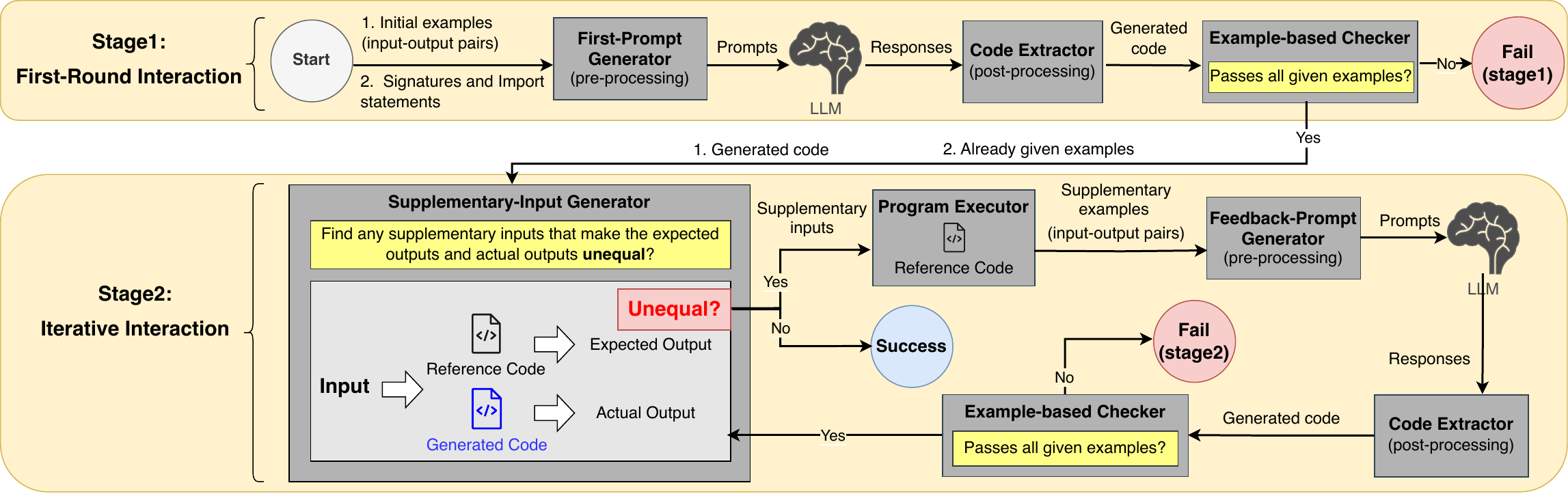}
\caption{The Iterative Evaluation Framework}
\label{fig: evaluating-framework}
\end{figure*}

Figure~\ref{fig: evaluating-framework} shows the overall workflow of the framework. 
The evaluation starts with the \textbf{first-round interaction} stage, where the prompts direct an LLM to generate code based on a set of given I/O examples. 
If the generated code cannot produce all the expected outputs for the given examples,
we consider the attempt to fail and end. 
In the \textbf{iterative interaction} stage, the framework iteratively checks whether the code successfully implements the target functionality, and supplements new I/O examples to clarify the discrepancies if it does not.
During the iterations, once the LLM generates code that conflicts with any given I/O example, we exit the iteration and regard it as a failed attempt.
Only if no I/O example can be found to demonstrate the discrepancies between the target and the actual functionalities, we consider the LLM to successfully implement the target functionality.

In both stages, the framework interacts with the LLM through two components: a prompt generator and a code extractor. The prompt generator receives I/O examples and the checking results of the generated code (if any), using them to construct prompts together with the signature of the target functionality.
The code extractor processes the LLM's answers, extracting all code snippets from the answers using regular-expression rules. It also filters the code that cannot be successfully compiled. The extracted code snippets are then sent to a code checker, which adopts different criteria in the first-round interaction and the iterative interaction. 

\textbf{Running Example.} To illustrate the evaluation framework, consider the programming task of ``checking whether the third integer equals the sum of the first two integers.'' Assume that the input parameters are integers ranging from 0 to 20, with each iteration providing three additional I/O examples. Initially, the set includes three I/O examples: (1) \CodeIn{function(1, 2, 3) = true}, (2) \CodeIn{function(10, 5, 2) = false}, and (3) \CodeIn{function(5, 2, 3) = false}.
In the first-round interaction, the three initial examples are directly used as tests for code checking. For instance, if the generated function returns \CodeIn{true} without doing anything, we consider it to fail because it does not yield the expected outputs for the last two examples.
However, code that passes the checking in first-round interaction may not always implement the target functionality, because these examples also specify many other functionalities (e.g., ``checking whether the sum of the first two integers is less than or equal to the third'', ``checking whether the three integers are increasing'', or even ``checking whether the first number is 1''). 
Therefore, it is necessary to supplement new I/O examples to clarify the target functionality in subsequent iterative interactions. 
For instance, if the generated code checks whether the three integers are increasing, the supplementary examples can be (1) \CodeIn{function(2, 3, 4) = false}, (2) \CodeIn{function(1, 2, 5) = false}, and (3) \CodeIn{function(1, 1, 2) = true}.

\subsection{First-Round Interaction}
The first-round interaction stage focuses on generating code for requirements in the form of only I/O examples.
In other words, we want to know whether an LLM can understand and implement the requirements described through only I/O examples. 
To achieve this purpose, the prompts inform the LLM to generate code according to the given I/O examples, and the code checker uses only the given I/O examples for testing.

\begin{figure*}[t]
\centering
\subfigure[The first prompt in first-round interaction stage]{
\begin{minipage}{0.49\textwidth}
\centering
\label{fig: prompt1}
\includegraphics[width=1\textwidth]{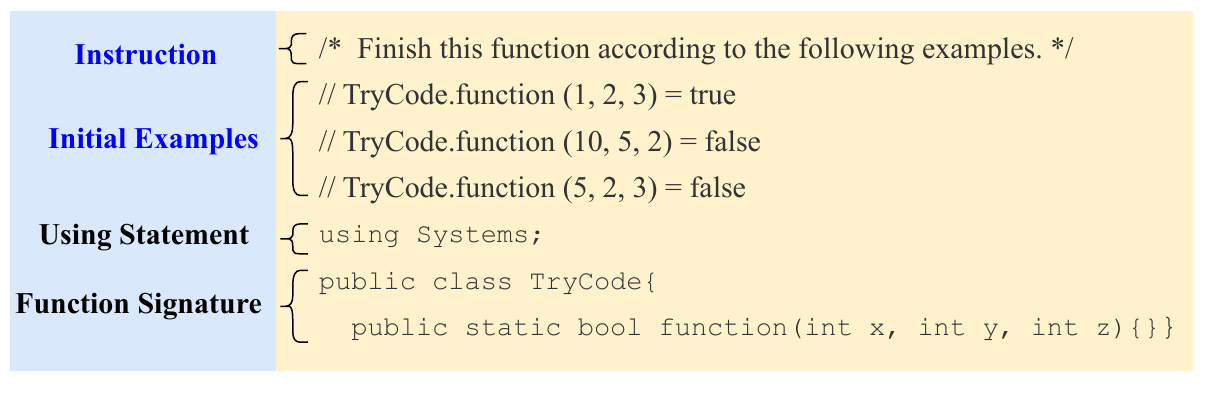}
\end{minipage}}
\subfigure[The feedback prompt in iterative interaction stage]{
\begin{minipage}{0.49\textwidth}
\centering
\label{fig: prompt2}
 \includegraphics[width=1\textwidth]{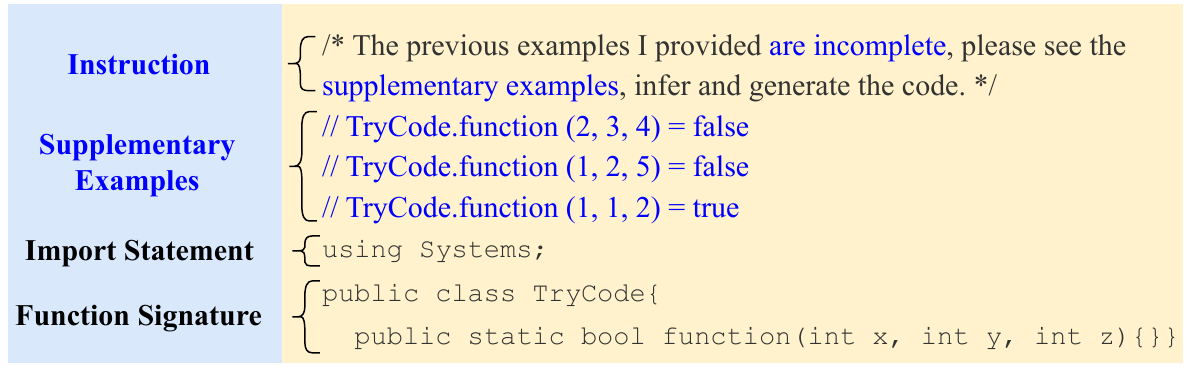}
\end{minipage}}
\caption{Instances of the Prompts in the Two Stages of the Evaluation Framework}
\label{Fig: motivation}
\end{figure*}

\textbf{Prompt Design.}
The prompt at this stage includes four parts: an instruction stating that I/O examples describe the requirement, a set of I/O examples, using statements, and the function signature.
For instance, the prompt for the running example in the first-round interaction is presented in Figure~\ref{fig: prompt1}.

\textbf{Code Checking.}
The code checking at this stage is implemented through unit testing, where each test case executes the generated function for one given input, collecting the actual output and comparing it with the expected one specified in the examples. Any discrepancy indicates a failure\footnote{Empirically, code with runtime errors (e.g., StackOverflow) or with a running time over 5000ms is also regarded as a failure} in the generation process.
After that, the code that passes all the unit tests (i.e., all given I/O examples) in this stage advances to the next iterative interaction stage.



\subsection{Iterative Interaction}
The iterative interaction stage focuses on implementing the target functionality from (iteratively supplemented) I/O examples.
For the target functionality, the framework requires reference code as the ground truth to create expected outputs of new examples. In each iteration, the framework checks whether the generated code is input-output equivalent to the reference code. If not, the framework adaptively provides new I/O examples and asks the LLM to modify its generated code. 

\textbf{Prompt Design.}
In addition to the information contained in the first prompt, the feedback prompt at this stage also introduces 
(1) the instruction that indicates that the previously provided I/O examples do not describe the functionality completely;
(2) the supplementary I/O examples where the previously generated code cannot produce the expected outputs.
For instance, the feedback prompt of the running example at this stage is presented in Figure~\ref{fig: prompt2}.
During the iterations, this new prompt is presented to the LLM along with the history of previous conversations.

\textbf{Code Checking.}
The code checking at this stage needs to address two questions: (1) whether the generated code conforms to all the given I/O examples and (2) whether the code implements the target functionality.
To answer the first question, the framework performs unit testing as it does during the first-round interaction, except that the number of test cases increases beyond the given examples. Only if the generated code passes all the unit tests does the checking move on.
To answer the second question, the framework tries to find supplementary inputs, which correspond to different outputs in the target functionality and the currently generated code.
Such inputs are generated in two ways: executing extensive unit testing and invoking a structural test generator. Extensive unit testing uses test cases constructed from a predefined broader set of I/O examples, which are currently invisible to the LLM.
Furthermore, the framework also employs a structural test generator~\cite{Klee, Dart, Cute, Pex} to check the generated code, aiming to reduce false positives caused by incomplete tests.
A structural test generator is typically guided by code coverage metrics~\cite{Klee}. To obtain supplementary inputs, the structural test generator tries to achieve full coverage of a wrapper function, which asserts that the generated code and the reference code have equal outputs.
If no supplementary inputs can be found, we approximately conclude that the generated code can always produce the same outputs as the target functionality, i.e., the LLM successfully implements the target functionality. Otherwise, we execute the reference code on each supplementary input to compose more comprehensive I/O examples, which are then presented to the LLM to clarify the target functionality.

%% file: NewBenchmark.tex
\section{\benchmark~ Benchmark}

We introduce \benchmark, our new benchmark designed specifically for evaluating example-based code generation. \benchmark~ consists of various programming tasks (i.e., target functionalities).
To adapt to the iterative evaluation framework, for each programming task, the benchmark should include three components:
(1) a ground-truth implementation (i.e., reference code) of the target functionality;
(2) a function signature indicating the input-output types; and
(3) input constraints specifying the range of the inputs.
In this section, we first list the sources and selection criteria for the programming tasks, and then describe the construction and usage of the three components.

\subsection{\stepone~} 
\benchmark~ includes programming tasks from two main sources.
(1) CodeHunt~\cite{CodeHunt, CodeHunt2}: a real-world dataset collected from a high-impact educational gaming platform where each task's requirement is presented with only tests (i.e., I/O examples). According to the given tests, the players need to iteratively modify their code (written in Java or C\#) to match the input-output behavior of an invisible secret function.
The programming tasks in CodeHunt are specifically designed for educational scenarios, including fundamental concepts such as control structures, data manipulation, and algorithmic problem solving. 
(2) HumanEval~\cite{Codex}, a well-known benchmark for code generation. It is a benchmark specifically tailored to Python, comprising a diverse set of programming tasks accompanied by natural-language statements, I/O examples, and test cases. HumanEval has been widely used~\cite{CodeGenerationSurvey} and extended~\cite{HumanevalExtension1, HumanevalExtension2} by existing studies, and many LLMs have achieved promising performance on this benchmark.
By involving the programming tasks from HumanEval, we can further compare LLMs' code generation capability between natural-language requirements and I/O-example requirements.

We manually filter the collected programming tasks based on their requirements, excluding those whose inputs and outputs cannot be represented by simple data structures (e.g., lists with elements of different data types). In total, we obtain 172 programming tasks for example-based code generation: 24 of these tasks are from CodeHunt, and 148 are from HumanEval.

\subsection{Construction}
To apply the original CodeHunt and HumanEval benchmark to our framework, it is necessary to (1) modify the standard answer to the chosen programming language, (2) assign appropriate function signatures, and (3) specify reasonable input constraints for the functions. Additionally, for clarity of description, we also simplify the input-output types or the functionalities for some programming tasks. 
Specifically, \benchmark~ and our evaluation use C\# as the programming language for code generation.

\textbf{Ground-Truth Implementation.}
As a recognized ``correct answer'' of the target functionality, the ground-truth implementation plays an important role in both code checking and supplementary example generation. 
CodeHunt already provides a C\# ground-truth implementation for each programming task. As for HumanEval, we manually prepare a C\# ground-truth implementation according to the functionality and the given answer written in Python.

\textbf{Function Signature.}
Function signatures, consisting of using statements, function names, and input-output types, are used to compose the prompts.
We make the using statements and input-output types consistent with the ground-truth implementation.
Different from most code-generation benchmarks, \benchmark~ sets a default function name (\emph{i.e., Puzzle}) for all programming tasks to avoid revealing the target functionalities through their function names.

\textbf{Input Constraints.} 
Explicit input constraints play an important role in iterative evaluation, also serving as a primary distinction among \benchmark~ and other code-generation benchmarks.
Input constraints are mainly used for input generation, including random generation and structural-test-generator-based generation (i.e., checking whether the code successfully implements the target functionality).
We ask experienced programmers to specify input constraints based on their understanding of each target functionality, ensuring that the I/O examples can be brief (fewer than 500 tokens each) and relevant, without compromising the accuracy of the description.\footnote{Some programming tasks restrict their input to meet complicated structures, which are difficult to represent explicitly through simple constraints. We manually write the input generators for these tasks.}

To cope with the possible bias introduced by random sampling, we prepare five sets of randomly sampled I/O examples for each programming task for the beginning of executions.
The number of I/O examples presented in each iteration is configurable.
Overall, we pre-sample a total of 5 $\times$ 10 random I/O examples, 
of which the currently model-invisible ones will be used for code checking in iterative interaction.
Once the generated code passes all unit tests constructed by the pre-sampled examples, we adopt a C\# structural test generator named Pex~\cite{Pex, Pex1, Pex2}, whose configuration by default is to generate tests with high block coverage, for further checking and supplementary-example generation.


%% file: Evaluation.tex
\section{Evaluation Setting}
This section presents the five research questions, introduces the evaluation metrics of each research question, describes the six evaluated LLMs and their model settings, and shows the results of preliminary experiments conducted for prompt design.

\subsection{Research Questions}
\textbf{RQ1: (Toward O1 in Section~\ref{sec:Introduction}) How effectively can the LLMs generate code conforming to all the given I/O examples?}
Unlike natural-language descriptions, I/O examples not only appear less frequently in training data, but also have higher diversity. This RQ aims to assess whether the LLMs can understand the requirements conveyed through I/O examples by checking whether the generated code conforms to all the given examples.

\textbf{RQ2: (Toward O2 in Section~\ref{sec:Introduction}) With the iteratively supplemented I/O examples, how effectively can the LLMs generate code for the target functionality?} The iteratively and adaptively supplemented I/O examples place higher demands on the LLMs' capability to generalize and understand.
Specifically, this RQ aims to evaluate the LLMs' capabilities to infer the target functionality from I/O examples and to improve the generated code through iterative feedback.

\textbf{RQ3: (Impact of natural language) How effectively can combining natural-language descriptions with I/O examples help improve the LLMs' score?} 
Although providing \emph{perfect} descriptions can be difficult, it is often feasible to provide related but imprecise information about the functionality. This RQ aims to explore the contribution of natural-language descriptions, especially those less precise, to example-based code generation for LLMs.

\textbf{RQ4: (Target-Functionality Analysis) What kinds of functionalities can be implemented through example-based code generation by the LLMs?}
The potential difficulty of code generation for different types of functionalities may vary. Particularly, for example-based code generation, this difficulty might also be determined by input-output types and the relevant knowledge of the functionality. 
This RQ aims to compare the LLMs' score across different types of functionalities and understand their strengths and weaknesses. The results may help us extend LLMs to suitable application scenarios.

\mr{\textbf{RQ5: (Impact of I/O examples) How much is the score of the LLMs affected by the selection of I/O examples? }
The difficulty of example-based code generation is related not only to the target functionality but also to the provided I/O examples.  
This RQ aims to assess the sensitivity of the LLMs' score to the choice of the given I/O examples, and the results may inspire future prompt engineering.}

\subsection{Metrics}
\begin{figure*}[t]
\centering
\includegraphics[width=0.9\textwidth]{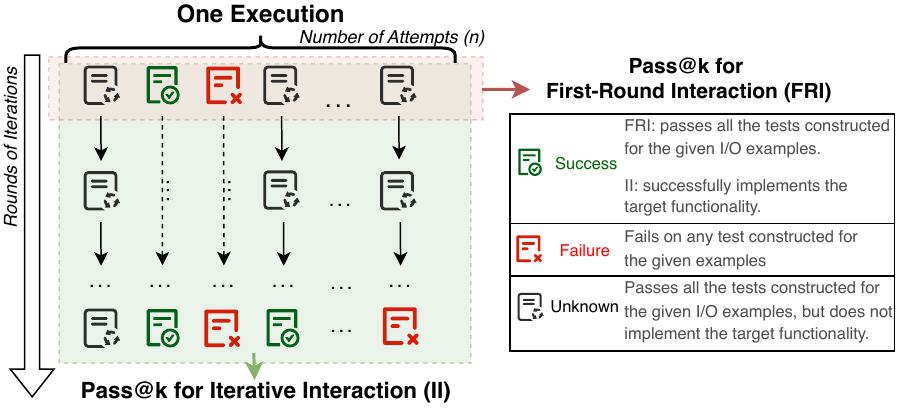}
\caption{Metrics of the Iterative Evaluation Framework}
\label{fig: metrics}
\vspace{-1em}
\end{figure*}

We adopt the \textbf{Pass@k} metric~\cite{Codex} (with $k$’s value of 1, 5, and 10, respectively), which measures the ability of an LLM to generate at least one correct code within $k$ attempts, as the primary metric to evaluate LLMs' capability on example-based code generation. 
Particularly, the decision of successful code generation differs in first-round interaction and iterative interaction. As shown in Figure~\ref{fig: metrics}, the evaluation starting from an initial set of I/O examples is called an $execution$.
For first-round interaction (RQ1), we set the total number of samples as 10 ($n=10$, i.e., making 10 attempts to generate code) in each execution. We consider the generated code in one attempt correct if it passes all the given tests constructed for the given I/O examples.
During the subsequent iterative interaction, we take the code that passes the first-round interaction as a starting point, asking the LLM to improve the code according to the conversation history and newly generated feedback (i.e., supplementary I/O examples), obtaining one answer each time. If an LLM finally succeeds in generating code for the target functionality during the iteration, we consider the attempt a successful one in the iterative interaction.

Additionally, to investigate how the number of I/O examples affects the correctness of code generation, we set the number of presented examples (Number of Examples, $NoE$) in each iteration to 3, 5, 7, and 10, respectively. Particularly, if the supplementary examples are generated through the structural test generator, we directly present all new examples to the LLM.

\subsection{Evaluated LLMs and Model Settings} 
We evaluate 
five state-of-the-art open-source LLMs (Gemma~\cite{gemmateam2024gemma}, CodeGemma~\cite{codegemma}, DeepSeek-Coder~\cite{DeepSeek-Coder}, Llama2~\cite{llama2}, and CodeLlama~\cite{CodeLlama}) and one close-source LLM (GPT-4o-mini~\cite{gpt4omini}). 

\begin{itemize}    

\item \textbf{DeepSeek-Coder~\cite{DeepSeek-Coder}}: A code generation model pre-trained from scratch on 2 trillion tokens of bilingual (English/Chinese) data, featuring an 87\% code and 13\% natural-language composition. The model is further instruction-tuned on 2 billion tokens of task-specific data.

\item \textbf{Gemma~\cite{gemmateam2024gemma} and CodeGemma~\cite{codegemma}}: The Gemma foundation model utilizes training data comprising English web documents, mathematical content, and code. Its code-specialized variant CodeGemma incorporates 500B-1T additional tokens of programming data and mathematical problem-solving content.

\item \textbf{Llama2~\cite{llama2} and CodeLlama~\cite{CodeLlama}}: Built upon the Llama2 architecture pre-trained on 2 trillion tokens of diverse open-source data, CodeLlama extends this foundation with 500B-1T tokens of domain-specific programming language data for code comprehension tasks.
    
\item \textbf{GPT-4o-mini~\cite{gpt4omini}}: GPT-4o-mini is a compact version of the GPT-4o model released by OpenAI. It is more intelligent than the previously widely evaluated GPT-3.5-turbo and is designed to achieve higher performance utilizing fewer computational resources. 
\end{itemize}

All six LLMs have been widely evaluated on code generation tasks~\cite{leaderboard}.
Additionally, the chosen open-source LLMs cover models obtained from three different training techniques: the base models, code models that are fine-tuned on code data from the base models, and models trained from scratch on code data.
For open-source LLMs, we evaluate their 7B (or nearly 7B) versions for three main purposes: (1) fairly comparing the performance of different models by eliminating the effect of model size,
(2) striking a balance between computational resources and model representativeness, and (3) making our experiments easy to reproduce even for researchers facing resource constraints.
Specifically, for DeepSeek-Coder, we choose deepseek-coder-6.7b-instruct\footnote{https://huggingface.co/deepseek-ai/deepseek-coder-6.7b-instruct} for evaluation;
for Gemma and CodeGemma, we choose gemma-7b-it\footnote{https://huggingface.co/google/gemma-7b-it} and codegemma-7b-it\footnote{https://huggingface.co/google/codegemma-7b-it}, respectively; for Llama2 and CodeLlama, we choose llama-2-7b-chat\footnote{https://huggingface.co/meta-llama/Llama-2-7b-chat-hf} and codellama-7b-instruct\footnote{https://huggingface.co/codellama/CodeLlama-7b-Instruct-hf}, respectively.

\textbf{Model Settings:}
For the open-source LLMs, we configure the data type to torch.bfloat16, \CodeIn{do\_sample = True}, \CodeIn{num\_return\_sequences=10}, and leave other parameters as recommended (e.g., temperature is by default set to 1.0).
For GPT-4o-mini, we call the official API and make other parameters consistent with the open-source LLMs.

\subsection{Prompt Design}
To determine appropriate prompts (which may have a non-negligible impact on LLMs' performance~\cite{prompt_infuence}), we first conduct a preliminary experiment with the five open-source LLMs on the 24 target functionalities drawn from CodeHunt.
The preliminary experiment compares the original manually designed prompt with three variants: COT (Chain of Thought)~\cite{COT}, Persona (assigning a specific role with its perspective to LLMs)~\cite{persona}, and Few-shot Learning~\cite{fewshot}.\footnote{The illustrations of each prompt variant and the results can be found in our project website~\cite{website}.}
We adopt the same model configuration as the subsequent experiments, except that only three I/O examples are provided in the prompts.
We find that the score ranking between the five LLMs is roughly the same under the different prompt variants. For sub-objective1 (i.e., conforming to the given I/O examples), the score under the original prompt is overall slightly lower than that under the Persona variant, but still better than that under the other two prompt variants. For sub-objective2 (i.e., implement the target functionality), the score under the original prompt is overall higher than that under all three other prompt variants.
The preliminary experiment illustrates that for iterative example-based code generation, simply adopting the three prompt variants cannot contribute to an obvious improvement for the LLMs. Therefore, we decide to use the manually designed original prompts in the subsequent experiments.

%% file: Results.tex
\section{Results and Analysis}
\label{sec: results}

\noindent \textbf{RQ1: How effectively can the LLMs generate code conforming to all the given I/O examples?}

\begin{table*}[t]
    \footnotesize
    \setlength{\tabcolsep}{2pt}
    \begin{center}
        \caption{The Average \emph{pass@k} of Each LLM in First-Round Interaction}
        \label{tab: passk_all_firstround}
        \begin{tabular}{|p{100pt}<{\centering}|p{20pt}<{\centering}|p{20pt}<{\centering}|p{20pt}<{\centering}|p{20pt}<{\centering}|p{20pt}<{\centering}|p{20pt}<{\centering}|p{20pt}<{\centering}|p{20pt}<{\centering}|p{20pt}<{\centering}|p{20pt}<{\centering}|p{20pt}<{\centering}|p{20pt}<{\centering}|}
        
            \hline
            \multirow{2}*{} & \multicolumn{3}{|c|}{$NoE$=3} & \multicolumn{3}{|c|}{$NoE$=5} & \multicolumn{3}{|c|}{$NoE$=7} & \multicolumn{3}{|c|}{$NoE$=10}\\
            \cline{2-13} 
            & k=1 & k=5 & k=10 & k=1 & k=5 & k=10 & k=1 & k=5 & k=10 & k=1 & k=5 & k=10 \\
            \hline
         GPT-4o-mini & \textbf{0.31} & \textbf{0.49} & \textbf{0.55} & \textbf{0.29} & \textbf{0.44} & \textbf{0.49} & \textbf{0.28} & \textbf{0.42} & \textbf{0.47} & \textbf{0.27} & \textbf{0.40} & \textbf{0.45}\\
         \hline
deepseek-coder-6.7b-instruct & \textbf{0.18} & \textbf{0.35} & 0.42 & \textbf{0.16} & \textbf{0.32} & \textbf{0.38} & \textbf{0.15} & \textbf{0.29} & \textbf{0.35} & \textbf{0.14} & \textbf{0.28} & \textbf{0.34}\\
\hline
gemma-7b-it & 0.11 & 0.21 & 0.25 & 0.09 & 0.18 & 0.21 & 0.07 & 0.15 & 0.18 & 0.06 & 0.14 & 0.17\\
codegemma-7b-it & 0.13 & 0.33 & \textbf{0.43} & 0.10 & 0.28 & 0.37 & 0.09 & 0.26 & \textbf{0.35} & 0.08 & 0.23 & 0.31\\
\hline
Llama-2-7b-chat & 0.09 & 0.26 & 0.33 & 0.07 & 0.21 & 0.29 & 0.06 & 0.18 & 0.26 & 0.05 & 0.15 & 0.21\\
CodeLlama-7b-Instruct & 0.10 & 0.26 & 0.33 & 0.09 & 0.22 & 0.29 & 0.08 & 0.21 & 0.27 & 0.07 & 0.18 & 0.23\\      
            \hline
            \end{tabular}
    \end{center}
\end{table*}
\begin{table*}[t]
    \footnotesize
    \setlength{\tabcolsep}{2pt}
    \begin{center}
        \caption{The Average \emph{pass@k} of Each LLM in Iterative Interaction}
        \label{tab: passk_during_iteration}
        \begin{tabular}{|p{100pt}<{\centering}|p{20pt}<{\centering}|p{20pt}<{\centering}|p{20pt}<{\centering}|p{20pt}<{\centering}|p{20pt}<{\centering}|p{20pt}<{\centering}|p{20pt}<{\centering}|p{20pt}<{\centering}|p{20pt}<{\centering}|p{20pt}<{\centering}|p{20pt}<{\centering}|p{20pt}<{\centering}|}
            \hline
            \multirow{2}*{} & \multicolumn{3}{|c|}{$NoE$=3} & \multicolumn{3}{|c|}{$NoE$=5} & \multicolumn{3}{|c|}{$NoE$=7} & \multicolumn{3}{|c|}{$NoE$=10}\\
            \cline{2-13} 
            & k=1 & k=5 & k=10 & k=1 & k=5 & k=10 & k=1 & k=5 & k=10 & k=1 & k=5 & k=10 \\
            \hline
            
GPT-4o-mini & \textbf{0.16} & \textbf{0.26} & \textbf{0.30} & \textbf{0.16} & \textbf{0.27} & \textbf{0.31} & \textbf{0.16} & \textbf{0.27} & \textbf{0.31} & \textbf{0.16} & \textbf{0.26} & \textbf{0.32}\\
\hline
deepseek-coder-6.7b-instruct & \textbf{0.09} & \textbf{0.18} & \textbf{0.22} & \textbf{0.10} & \textbf{0.19} & \textbf{0.23} & \textbf{0.10} & \textbf{0.19} & \textbf{0.23} & \textbf{0.10} & \textbf{0.20} & \textbf{0.24}\\
\hline
gemma-7b-it & 0.03 & 0.05 & 0.07 & 0.03 & 0.06 & 0.07 & 0.03 & 0.05 & 0.06 & 0.03 & 0.05 & 0.07\\
codegemma-7b-it & 0.02 & 0.07 & 0.11 & 0.03 & 0.08 & 0.12 & 0.03 & 0.08 & 0.12 & 0.03 & 0.08 & 0.11\\
\hline
Llama-2-7b-chat & 0.01 & 0.02 & 0.03 & 0.01 & 0.02 & 0.04 & 0.00 & 0.02 & 0.04 & 0.00 & 0.02 & s0.03\\
CodeLlama-7b-Instruct & 0.03 & 0.08 & 0.12 & 0.04 & 0.09 & 0.13 & 0.03 & 0.10 & 0.13 & 0.03 & 0.09 & 0.12\\
\hline
            \end{tabular}
    \end{center}
\end{table*}

Considering only the tests constructed for the given I/O examples, the average pass@k of each LLM in first-round interaction is presented in Table~\ref{tab: passk_all_firstround}, where the best numbers among all the LLMs and those among the open-source LLMs are bolded. 
As the results show, all the LLMs demonstrate their capabilities, which still have room for improvement, in generating code conforming to all the given I/O examples. GPT-4o-mini outperforms all the evaluated open-source LLMs, achieving an average pass@10 ranging from 0.45 to 0.55. In contrast, the average pass@10 score of all the evaluated open-source LLMs is below 0.45, and as for pass@1, the average score sometimes even goes below 0.1. 
At the same time, the average pass@k score of all LLMs decreases as the number of the given I/O examples increases. 


\begin{figure*}
\centering
\lstset{
    language = Java,
    tabsize=2,
    basicstyle = \footnotesize\ttfamily,
    commentstyle =\color{gray!100},
    numberstyle = \small, 
    breaklines = true,    
    numbers = left,   
    keywordstyle = \color{blue},            
    stringstyle = \color{red!100},          
    showspaces = false,     
    columns = fixed,
    escapeinside = ``,
    linewidth = \linewidth,
    numbersep=5pt,
    xleftmargin = 3em
}
\newsavebox\mybubble
\begin{lrbox}{\mybubble}
\begin{lstlisting}
// TryCode.Puzzle(17) = 34
// TryCode.Puzzle(35) = 52
// TryCode.Puzzle(-21) = -4
using System.Collections.Generic;
using System;

public class TryCode
{
    public static int Puzzle(int x)
    {
        return (x * x - 10) % 100;
    }
}
\end{lstlisting}
\end{lrbox}

\newsavebox\myselection
\begin{lrbox}{\myselection}
\begin{lstlisting}
// TryCode.Puzzle(17) = 34
// TryCode.Puzzle(35) = 52
// TryCode.Puzzle(-21) = -4
using System;
using System.Collections.Generic;
public class TryCode{
    public static int Puzzle(int x){
        if (x == 17) return 34;
        if (x == 35) return 52;
        if (x == -21) return -4;
        return 0; // default return
    }
}
\end{lstlisting}
\end{lrbox}

\subfigure[An Instance of Code Generated by CodeLlama]{
\begin{minipage}[b]{0.45\linewidth}
\usebox\mybubble \\
\end{minipage}
\label{Fig: CodeLlama}
}
\subfigure[An Instance of Code Generated by Llama-2]{
\begin{minipage}[b]{0.45\linewidth}
\usebox\myselection \\
\end{minipage}
\label{Fig: Llama}
}
\caption{Different Programming ``Preference'' of Llama-2-7b-chat and CodeLlama-7b-Instruct}
\vspace{-1em}
\label{Fig: PreferenceExample}

\end{figure*} 

Surprisingly, additional code-related training does not always obviously improve performance in our evaluation. Codegemma-7b-it substantially surpasses its base model (gemma-7b-it), and yet CodeLlama-7b-Instruct's score is very close to that of its base, Llama-2-7b-chat. Analyzing the 15 cases where Llama-2-7b-chat most outperforms CodeLlama-7b-Instruct reveals differing code generation ``preferences''.  Llama-2-7b-chat tends to perform \emph{input matching}, directly generating conditional statements from the I/O examples without inferring input-output relationships. Conversely, CodeLlama-7b-Instruct attempts to deduce these relationships, but these deductions often fail to satisfy even the provided I/O examples. Figure \ref{Fig: PreferenceExample} illustrates this observation: Llama-2-7b-chat uses separate \CodeIn{if} statements to match each I/O example, while CodeLlama-7b-Instruct incorrectly proposes an arithmetic expression (intended to be \CodeIn{x+17}) that fails all the given examples.
 
To identify the effect of \emph{input-matching} behavior on the results, we use string checking to filter the code that performs \emph{input matching}, finding that all the LLMs may generate code performing \emph{input matching}. The largest proportion of such code is found in that generated by Llama-2-7b-chat, while in the code generated by deepseek-coder-6.7b-instruct and gemma-7b-it, the percentage is considerably smaller.

\begin{tcolorbox}[colback=lightgray, colframe=gray, title=\textbf{Summary of RQ1}]
\noindent \textbf{Overall Assessment:} 
All the LLMs struggle to consistently generate code that conforms to all the given I/O examples. Moreover, the score of code generation decreases as the number of the given examples increases.

\noindent \textbf{Model Comparison:} 
GPT-4o-mini outperforms all the open-source LLMs, and deepseek-coder-6.7b-instruct leads among the open-source LLMs.

\noindent \textbf{Special-Case Analysis:} Given I/O examples, all the LLMs may generate code that simply performs \emph{input matching}. Llama-2-7b-chat is the most severe over the LLMs.
\end{tcolorbox}


\noindent \textbf{RQ2: With the iteratively supplemented I/O examples, how effectively can the LLMs generate
code for the target functionality?}

\begin{table*}[]
\footnotesize
\caption{The Average \emph{pass@k} of Each LLM When Given Natural-Language Descriptions and Human-Designed Examples}
\label{tab: NLscore}
\begin{tabular}{|c|r|r|r|c|r|r|r|}
\hline
LLM & \multicolumn{1}{l|}{k=1} & \multicolumn{1}{l|}{k=5} & \multicolumn{1}{l|}{k=10} & LLM & \multicolumn{1}{l|}{k=1} & \multicolumn{1}{l|}{k=5} & \multicolumn{1}{l|}{k=10} \\ \hline
GPT-4o-mini& 0.84& 0.90& 0.90& codegemma-7b-it & 0.47& 0.73& 0.80\\
deepseek-coder-6.7b-instruct & 0.65& 0.86& 0.88& Llama-2-7b-chat-hf    & 0.19& 0.29& 0.33\\
gemma-7b-it& 0.29& 0.45& 0.52& CodeLlama-7b-Instruct & 0.33& 0.61& 0.69\\ \hline
\end{tabular}
\end{table*}

\begin{figure*}[t]
\centering
\includegraphics[width=0.98\textwidth]{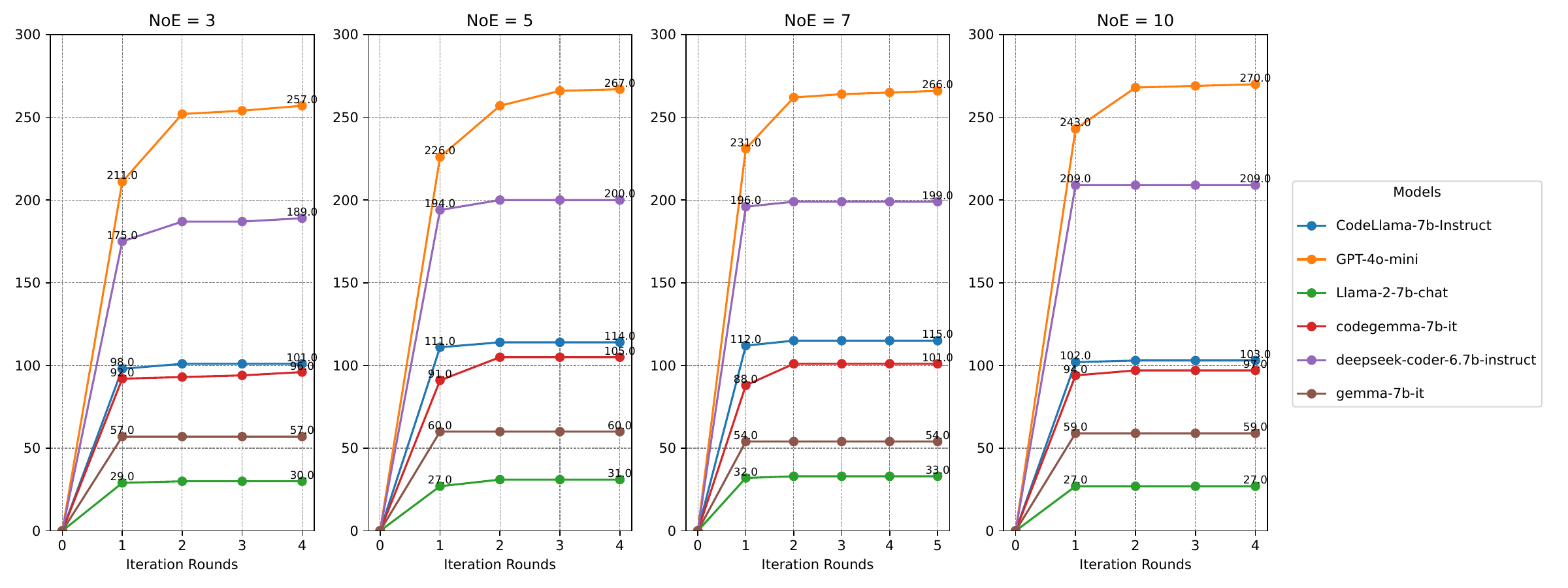}
\caption{The Cumulative Number of Successful Executions during Iterative Interaction}
\label{Fig: successful executions during iteration}
\end{figure*}

Table \ref{tab: passk_during_iteration} shows the pass@k of all the LLMs within five rounds of iteration, where the best numbers among all the LLMs and those among the open-source LLMs are bolded. 
For comparison, we also evaluate these LLMs when the target functionality is described through its original prompts (including natural-language description and some human-designed I/O examples) and the score is presented in Table~\ref{tab: NLscore}.
According to the results,
we can see that the pass@k score no longer decreases monotonically with the number of the given examples in each round.
Compared to the score of executions where natural-language descriptions are given, all the LLMs experience an obvious decline. Even the average pass@1 score of GPT-4o-mini is less than 0.2.
Particularly, the pass@1 score of Llama-2-7b-chat is approaching 0, illustrating that it is difficult for this LLM to correctly infer functionalities from I/O examples.
The ranking among all the LLMs is roughly consistent with that in RQ1, but the distinction among the open-source LLMs becomes higher.
GPT-4o-mini is still the best, outperforming all the open-source LLMs. deepseek-coder-6.7b-instruct is the best of all the open-source LLMs, followed by codegemma-7b-it and CodeLlama-7b-Instruct.
Interestingly, the advantage of CodeLlama-7b-Instruct over Llama-2-7b-chat is markedly greater than that of codegemma-7b-it over gemma-7b-it.

To visualize the change in score during the iteration, Figure~\ref{Fig: successful executions during iteration} counts the number of executions in which the target functionality has been successfully implemented after each round of iteration, out of a total of $172 \times 5 = 860$ executions.
First, we can see that the evaluated LLMs exhibit obvious differences after the first iteration, indicating their \textbf{varying capabilities in inferring the target functionality from I/O examples}: GPT-4o-mini performs the best, deepseek-coder-6.7b-instruct is superior to other open-source LLMs, and Llama-2-7b-chat achieves the least success.
Second, it is also worth noting that all the LLMs progress most in the initial one or two rounds, indicating that in most cases, these LLMs \textbf{can hardly successfully improve the code according to iterative feedback}. Specifically, for GPT-4o-mini, about ninety percent of the executions succeed within the first two rounds of iteration. As for most of the open-source LLMs, the iteratively supplemented I/O examples (after the first round) contribute to new success on no more than 10 executions. We also analyze the proportion of errors occurring in all executions. The most frequent error is ``fail in the given tests'', indicating that as the number of iteration rounds increases, the LLMs struggle to generate code that satisfies all the given I/O examples.

\begin{tcolorbox}[colback=lightgray, colframe=gray,  title=\textbf{Summary of RQ2}]
\noindent \textbf{Overall Assessment:} Generating code with (iteratively supplemented) I/O examples remains a challenging task for all the evaluated LLMs. Even for those successful executions, the majority of successful code generation is achieved in the first round of the iterative process.

\noindent \textbf{Model Comparison:} GPT-4o-mini still outperforms all the evaluated LLMs. 
Deepseek-coder-6.7b-instruct outperforms other open-source LLMs, benefiting mainly from its superior capability in ``inferring the target functionality from I/O examples''.

\end{tcolorbox}

\noindent \textbf{RQ3: How effectively can combining natural-language descriptions with I/O examples help improve the LLMs' score?}

To evaluate the influence of natural-language descriptions, we evaluate the LLMs on prompts combining I/O examples with two different granularities (i.e., full descriptions and keywords) of natural-language descriptions, and then calculate their pass@5 score. Unlike full descriptions, the keywords for each target functionality point out only the general direction but do not tell the details (e.g., for the full description ``return x+17'', the keyword is ``Arithmetic Operation''). The results are given in Table~\ref{tab: NLComparison}, where the best numbers among all the LLMs and those among the open-source LLMs are bolded.

From Table~\ref{tab: NLComparison}, we can see that combining natural-language descriptions with I/O examples can greatly improve the LLMs' score, and the improvement is more pronounced in iterative interaction than in first-round interaction. At the same time, even without precise descriptions, relevant keywords of the functionality can still lead to considerable improvement. For open-source LLMs in iteration interaction, including keywords in the prompts can even help double the score.

\begin{tcolorbox}[colback=lightgray, colframe=gray,  title=\textbf{Summary of RQ3}]
\noindent Combining natural-language descriptions with I/O examples can greatly improve the LLMs' performance. Furthermore, when precise descriptions are unavailable, even related keywords can also enhance the score.
\end{tcolorbox}

\begin{table*}[]
    \footnotesize
    \setlength{\tabcolsep}{2pt}
        \begin{center}
        \caption{\mr{The Comparison between Combining I/O Examples with Different Granularity of Natural Language}}
        \label{tab: NLComparison}
            
            \begin{tabular}{|c|ccc|ccc|}
            \hline
            \multirow{2}{*}{} & \multicolumn{3}{c|}{\textbf{First-Round Interaction}}                        & \multicolumn{3}{c|}{\textbf{Iterative Interaction}}                        \\ \cline{2-7} 
             &
              \multicolumn{1}{l|}{\textbf{Only I/O}} &
              \multicolumn{1}{l|}{\textbf{I/O $+$ Keywords}} &
              \multicolumn{1}{l|}{\textbf{I/O $+$ full NL}} &
              \multicolumn{1}{l|}{\textbf{Only I/O}} &
              \multicolumn{1}{l|}{\textbf{I/O $+$ Keywords}} &
              \multicolumn{1}{l|}{\textbf{I/O $+$ full NL}} \\ \hline
            GPT-4o-mini &
              \multicolumn{1}{c|}{\textbf{0.49}} &
              \multicolumn{1}{c|}{\textbf{0.61}} &
              \textbf{0.92} &
              \multicolumn{1}{c|}{\textbf{0.26}} &
              \multicolumn{1}{c|}{\textbf{0.45}} &
              \textbf{0.88} \\ \hline
            deepseek-coder-6.7b-instruct &
              \multicolumn{1}{c|}{\textbf{0.35}} &
              \multicolumn{1}{c|}{\textbf{0.51}} &
              \textbf{0.88} &
              \multicolumn{1}{c|}{\textbf{0.18}} &
              \multicolumn{1}{c|}{\textbf{0.37}} &
              \textbf{0.83} \\ \hline
            gemma-7b-it                          & \multicolumn{1}{c|}{0.21} & \multicolumn{1}{c|}{0.25} & 0.49 & \multicolumn{1}{c|}{0.05} & \multicolumn{1}{c|}{0.11} & 0.32 \\ 
            codegemma-7b-it                      & \multicolumn{1}{c|}{0.33} & \multicolumn{1}{c|}{0.45} & 0.80 & \multicolumn{1}{c|}{0.07} & \multicolumn{1}{c|}{0.24} & 0.69 \\ \hline
            Llama-2-7b-chat-hf                   & \multicolumn{1}{c|}{0.26} & \multicolumn{1}{c|}{0.30} & 0.41 & \multicolumn{1}{c|}{0.02} & \multicolumn{1}{c|}{0.09} & 0.26 \\ 
            CodeLlama-7b-Instruct                & \multicolumn{1}{c|}{0.26} & \multicolumn{1}{c|}{0.37} & 0.69 & \multicolumn{1}{c|}{0.08} & \multicolumn{1}{c|}{0.20}  & 0.57 \\ \hline
            \end{tabular}

  \end{center}
  \vspace{-1em}
\end{table*}

\noindent \textbf{RQ4: What kinds of functionalities can be implemented through example-based code generation by the LLMs?}

To answer RQ4, we categorize the target functionalities from three dimensions: programming difficulty, input-output types, and related knowledge.
First, to categorize the difficulty of functionalities, we take ``whether a specific functionality can be successfully implemented given a natural-language description'' as the reference, yielding two categories: NL-Succeed and NL-Fail. 
Specifically, given the natural-language descriptions, if a model correctly implements the functionality in at least one out of ten generations, the functionality is categorized as NL-Succeed. Otherwise, it is categorized as NL-Failure.
It is important to note that due to the varying capabilities of LLMs, the categorization regarding difficulties also varies across different LLMs.
Second, the input-output types of functionalities are determined by the function signature, including five categories: int, string, double, array (with elements of any type), and boolean. Because each functionality may have multiple inputs, it may be categorized into multiple possible categories as well.
Third, the relevant knowledge of functionalities is manually labeled, and each functionality may correspond to multiple labels.

To visualize the results, we use horizontal bar charts to present the success rate in each category. 
For one functionality, if a model succeeds in any of the five executions, we regard the functionality as successfully implemented.
The length of the bar reflects the number of target functionalities, and the shaded portion indicates those successfully implemented from I/O examples. The proportion of successfully implemented functionalities is represented by the numbers adjacent to the shaded portion.

\begin{figure*}[t]
\centering
\includegraphics[width=0.8\textwidth]{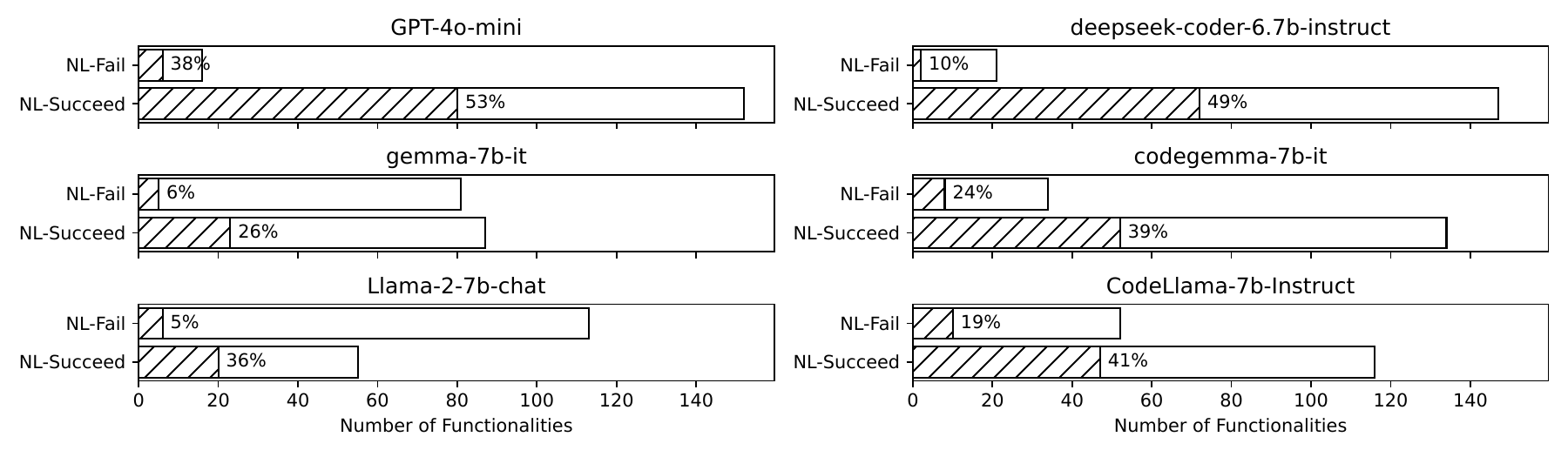}
\caption{The Success Rate of Functionalities with Different Difficulties}
\label{fig: success_rate_difficulty}
\end{figure*}

\begin{figure*}[t]
\centering
\includegraphics[width=0.8\textwidth]{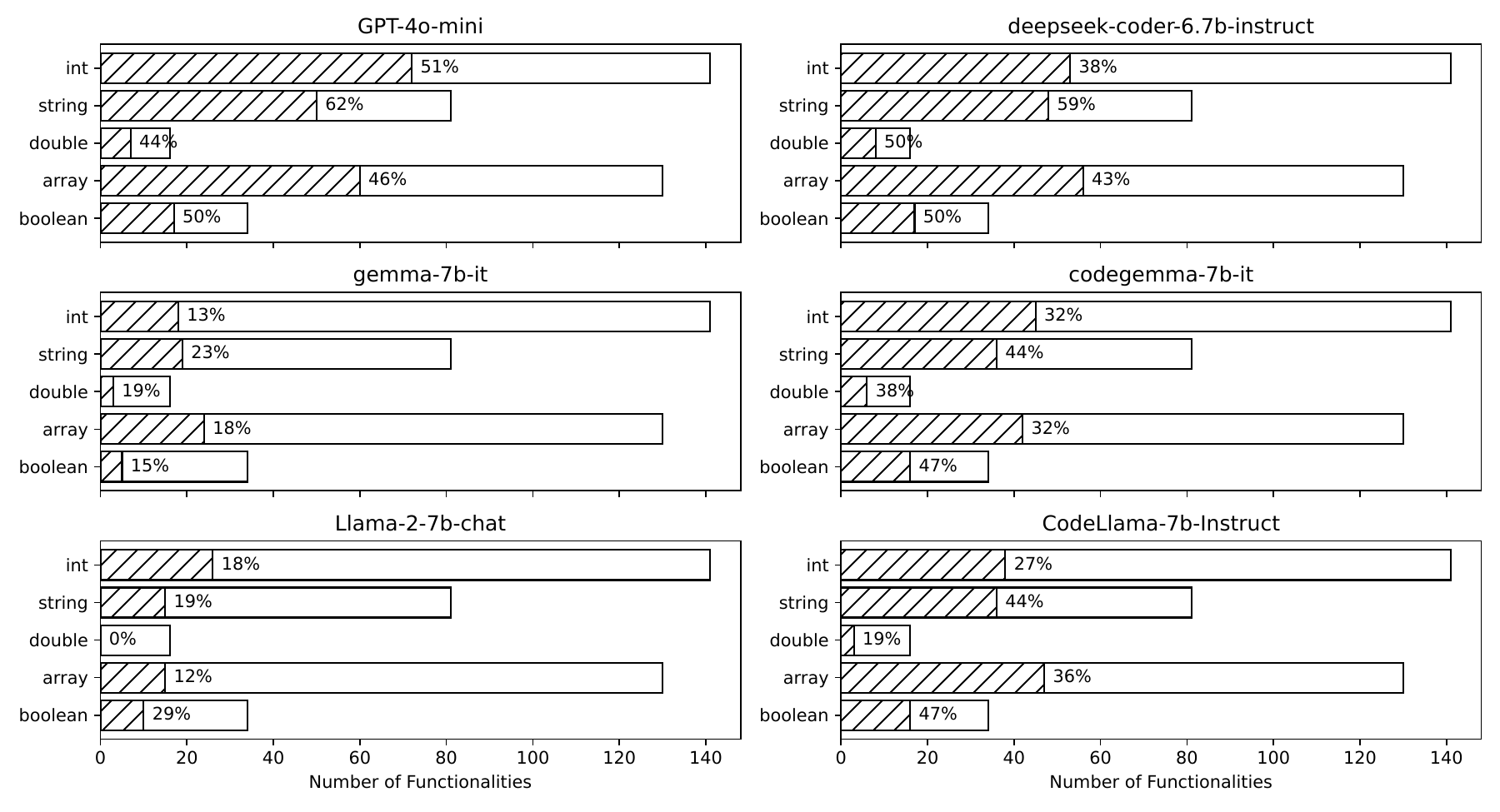}
\caption{The Success Rate of Functionalities with Different Input-Output Types}
\label{fig: success_rate_iotype}
\end{figure*}

\begin{figure*}[t]
\centering
\includegraphics[width=0.85\textwidth]{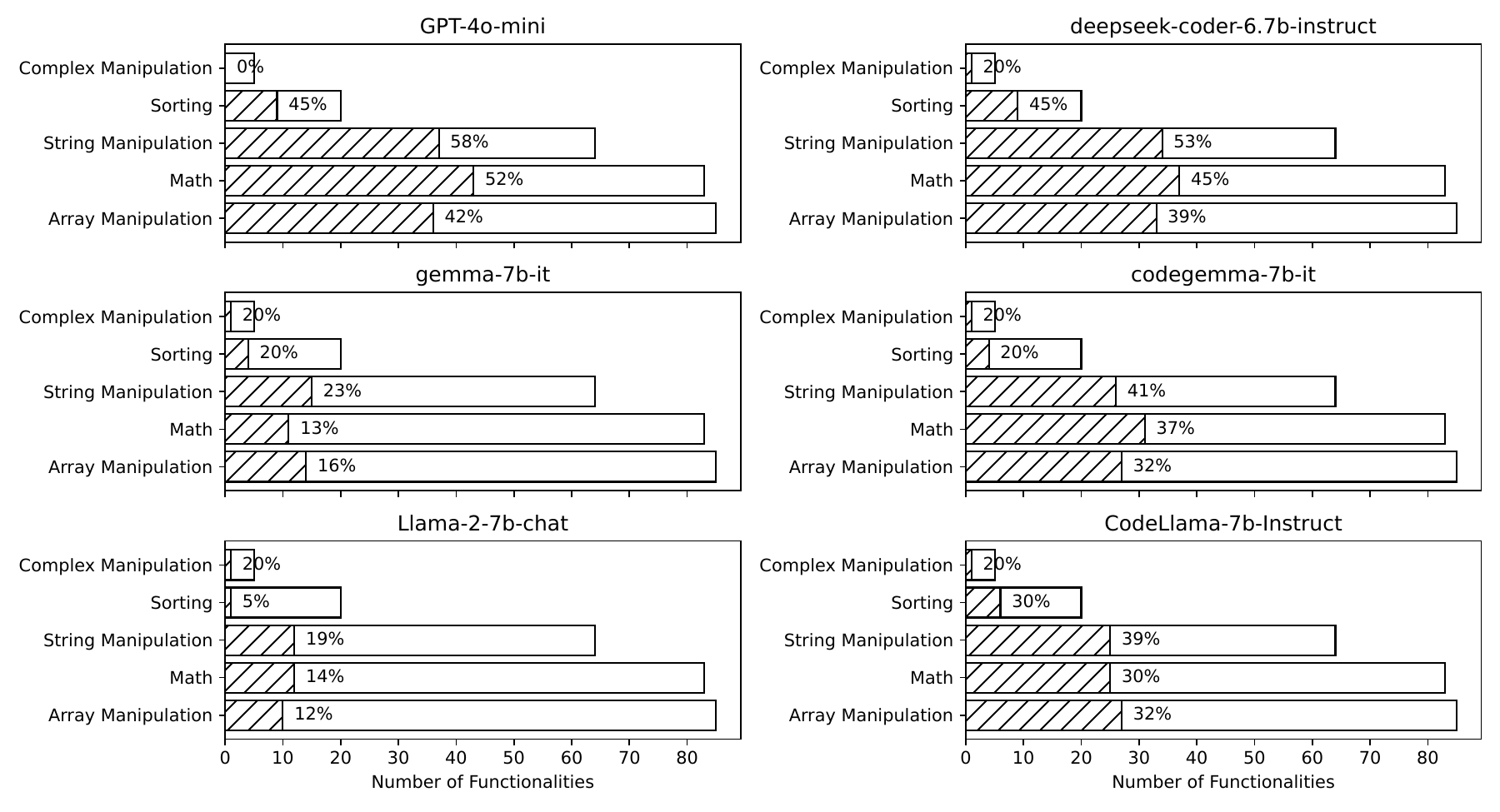}
\vspace{-1em}
\caption{The Success Rate of Functionalities with Different Related Knowledge}
\label{fig: success_rate_knowledge}
\end{figure*}

The success rate of functionalities with \textbf{different difficulties} is shown in Figure~\ref{fig: success_rate_difficulty}.
For all the evaluated LLMs, the majority of functionalities that can be implemented from I/O examples fall within the NL-Succeed category, but not all functionalities within this category can be implemented according to I/O examples. 
Even for the best model GPT-4o-mini, more than 40\% of the code in the NL-Succeed category cannot be implemented using I/O examples. 
We also observe that although the number of functionalities in deepseek-coder-6.7b-instruct's NL-Succeed category is close to that of GPT-4o-mini, the success rate of example-based code generation in this category is much lower than that of GPT-4o-mini.
Overall, the results confirm our hypothesis that generating code from I/O examples is a more challenging task than generating code from natural-language descriptions.

Additionally, functionalities that cannot be implemented based on natural-language descriptions but can be implemented according to I/O examples catch our interest. By manually inspecting these functionalities, we find that most of them (e.g., HumanEval-41) have the following characteristics: the natural-language descriptions introduce additional information that makes it even difficult to comprehend, but the relationships between inputs and outputs are straightforward or enumerable. 

The success rate of functionalities with \textbf{different input-output types} is shown in Figure~\ref{fig: success_rate_iotype}. We find that GPT-4o-mini, deepseek-coder-6.7b-instruct, and gemma-7b-it achieve their best results on functionalities with string-type inputs or outputs, and the other three LLMs achieve their best results with boolean types.
Particularly, despite being inferior to GPT-4o-mini overall, deepseek-coder-6.7b-instruct outperforms GPT-4o-mini in the category of functionalities involving double-type inputs or outputs.

The success rate of functionalities with \textbf{different related knowledge} is shown in Figure~\ref{fig: success_rate_knowledge}.
We find that all the LLMs exhibit the highest success rate in functionalities related to string manipulation. 
The second-highest success rate falls in categories related to math and array manipulation.
Particularly, for one of the functionalities related to complex manipulation (i.e., manipulations of complex data structures such as trees, graphs, and hash tables), the open-source LLMs outperform GPT-4o-mini.



\begin{tcolorbox}[colback=lightgray, colframe=gray,  title=\textbf{Summary of RQ4}]
\noindent Regarding programming difficulty, the LLMs tend to achieve higher success rates on functionalities that (1) can be successfully implemented according to natural-language descriptions, or (2) have an easy-to-catch relationship between the input and output.
Regarding input-output types, the LLMs perform the best on functionalities with string-type or boolean-type inputs and outputs.
Regarding related knowledge, the evaluated LLMs achieve their highest success rates on functionalities related to string manipulation.
\end{tcolorbox}

\noindent \textbf{RQ5: How much is the score of LLMs affected by the selection of I/O examples?}
\begin{figure*}[t]
\centering
\subfigure[Score Variance on Different Sets of I/O Examples]{
\begin{minipage}{0.49\textwidth}
\centering
\label{fig: variance0}
\includegraphics[width=1\textwidth]{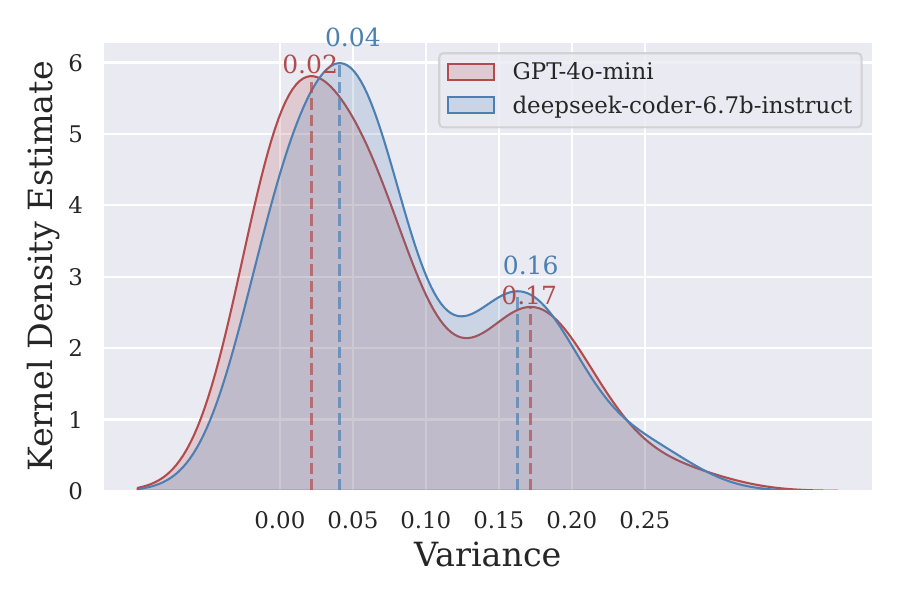}
\end{minipage}}
\subfigure[Impact Values of Common Examples]{
\begin{minipage}{0.49\textwidth}
\centering
\label{fig: variance1}
 \includegraphics[width=1\textwidth]{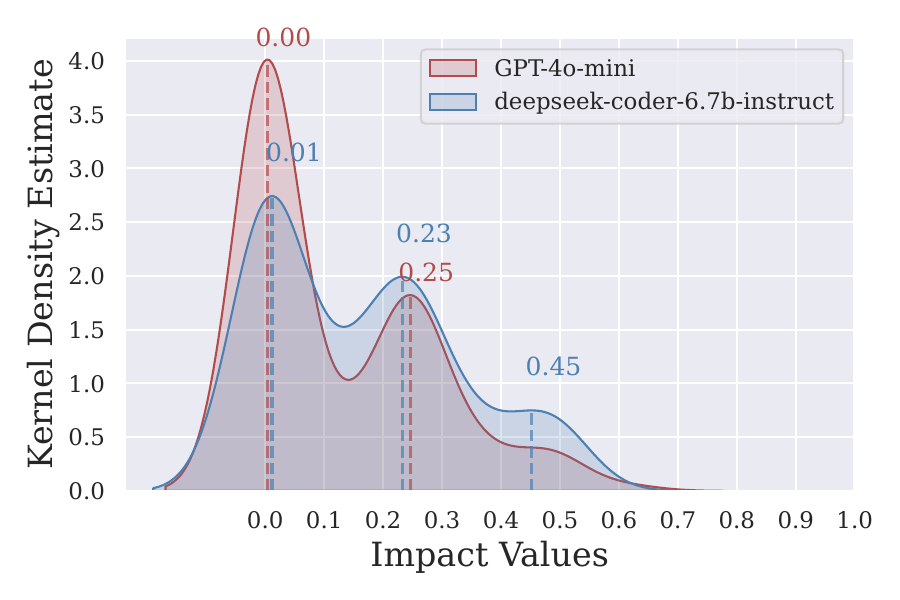}
\end{minipage}}
\caption{Analysis on the Impact of Different I/O Examples}
\label{Fig: motivation}
\end{figure*}

According to the conclusions from RQ2, the first prompt plays a crucial role in this iterative code generation process. Therefore, the analysis of this RQ mainly focuses on the impact of the examples provided in the first round.
Specifically, for the best-performing GPT-4o-mini and deepseek-coder-6.7b-instruct, we collect their pass@5 score of the five executions (starting from different sets of I/O examples, with 10 attempts in each execution) for the same target functionalities, and calculate their variance. Note that if an LLM does not succeed in any of the five executions, the corresponding target functionality is not included.

The distribution of the variance is presented by a kernel density estimate plot in Figure \ref{fig: variance0}. 
We observe ‌a bimodal structure for both deepseek-coder-6.7b-instruct and GPT-4o-mini. 
The primary peak of the variance density falls between $(0.00, 0.05)$, indicating that for a large portion of the functionalities, the two LLMs both have close pass@5 scores on different I/O-example sets. 
We also notice some functionalities for which the variance lies between $(0.15, 0.2)$ (corresponding to the secondary peak). For these functionalities, providing \emph{appropriate} I/O examples in the prompts may be helpful for code generation.

We try to identify the single I/O examples with high impact on the score. 
Specifically, for each I/O example appearing in more than one initial example set, we collect the average score when the example \emph{is} and \emph{is not} included, respectively, and calculate the difference between the scores. 
This difference, serving as the \emph{impact value}, is used to quantify the impact of a particular I/O example on the code-generation score, and its distribution is shown in Figure~\ref{fig: variance1}. A higher impact value implies that the I/O example has a greater impact on the final score. According to the figure, we find that the impact values of I/O examples show similar distributions for GPT-4o-mini and deepseek-coder-6.7b-instruct. Although most I/O examples have impact values close to 0, there are still a considerable number of I/O examples with impact values around 0.25. Particularly, our analysis also identifies about 10 I/O examples with impact values greater than 0.4.
Identifying such I/O examples with high impact value in advance may be an important direction for future improvements through prompt engineering.

\begin{tcolorbox}[colback=lightgray, colframe=gray,  title=\textbf{Summary of RQ5}]
\noindent In addition to functionality-related characteristics, the LLMs' score is also affected by the selection of I/O examples provided in prompts.
Including high-impact examples may bring an improvement to the score.
\end{tcolorbox}


%% file: Recommendation.tex
\section{Recommendation}
\mr{This section presents recommendations for enhancing LLM-driven code generation according to the evaluation results obtained in our work. For LLM developers, we suggest directions to improve LLMs for code generation; for users, we provide strategies to better utilize LLMs for code generation.}

\subsection{Directions for Improving LLMs}
\textbf{Supporting diverse formats of requirements.} Current research on LLMs for code generation tasks uses natural language as the main format to describe the target functionalities. However, suitable natural-language descriptions are not always available in real life, because the target functionalities can be unknown (e.g., for reverse engineering tasks) or difficult to describe clearly (e.g., for end-users without sufficient programming knowledge). In this study, we describe the functionalities mainly by I/O examples, which are unambiguous and easily accessible, finding that LLMs still struggle to generate code conforming to the given I/O examples. This finding suggests that LLMs currently do not support the I/O-example-form of the descriptions very well. Therefore, we propose to focus more on and improve LLMs' code-generation capability in other forms of requirement description besides natural language. 
Doing so can help not only comprehensively understand the capability boundary of LLMs but also extend LLM-based code generation to more application scenarios.

\textbf{Supporting multi-turn and iteratively given requirements.}
It is not rare to see that single-turn given requirements are not sufficient to describe the target functionalities completely. This study, which provides I/O examples for the target functionalities, faces the same situation. However, after introducing iteratively supplemented I/O examples, we find that the LLMs can hardly effectively utilize them. In most cases, the generated code either fails to simultaneously satisfy the examples given in multiple turns, or is influenced by past answers, simply adding special cases to adapt to new I/O examples. Therefore, we suggest paying attention to LLMs' code-generation capability with multi-turn requirements, especially the capability to integrate multi-turn requirements and effectively utilize feedback.

\subsection{Strategies to Better Utilize LLMs}
The findings of this study also shed light on the usage strategies for users hoping to directly utilize LLMs for programming tasks. These strategies are also useful for other similar tasks that may need multi-turn given requirements.

\textbf{Valuing early prompts.}
Our evaluation of LLMs reveals that in iterative example-based code generation, early (especially the first) prompts play a crucial role in ultimately implementing the target functionality correctly. Therefore, when applying LLMs to iterative example-based code generation, we should emphasize the design of the first prompt, instead of relying too much on making supplements and corrections in subsequent interactions. One possible idea is to choose I/O examples that are as \emph{representative} (whose definition remains to be explored) as possible in the first prompt while adding special cases in subsequent prompts.

\textbf{Rebooting timely.}
Considering that LLMs may not be able to effectively utilize the multi-turn given requirements, when the number of iteration rounds increases to a certain threshold, restarting the conversation may be a better option than continuing with iterations. When restarting, the prompts provided in the completed iterations need to be re-selected and recombined to construct the new initial prompt, aiming to improve the accuracy of code generation.

\textbf{Providing relevant natural-language information (even being inaccurate).}
The evaluation demonstrates that combining I/O examples with domain-related keywords (i.e., terms indicating functionality directions without detailed steps) can improve LLMs' performance. Therefore, when good natural-language descriptions are not available, a feasible improvement is to provide the LLM with keywords related to the target functionalities. Even if these keywords are not accurate, they are still likely to be of great help.

%% file: Related.tex
\section{Related Work}
The significant progress of LLMs in code generation has also propelled the research to evaluate their capabilities. Recently, many efforts have been made to evaluate LLMs from different programming languages, different difficulty levels, and different applications.

\textbf{Benchmarks of different programming languages.} Benchmarks on Python code~\cite{Codex, MBPP, APPS, DS-1000} make up a large part of existing efforts. In addition to Python, researchers have also constructed benchmarks for other widely used programming languages (e.g., AixBench~\cite{AixBench} for Java) and domain-specific languages (e.g., BIRD~\cite{BIRD} for SQL). Multiple-E~\cite{MultiPL-E} includes programs written by 18 programming languages in addition to Python.
    
\textbf{Benchmarks of different difficulty levels.} Existing benchmarks consider programming problems from entry-level to industrial-level. MBPP~\cite{MBPP} is designed with hundreds of entry-level problems, e.g., numeric manipulations. HumanEval~\cite{Codex} includes 164 human-written programming problems from introductory to interview style and is relatively easy. APPS~\cite{APPS}, CodeContests~\cite{AlphaCode}, TACO~\cite{TACO} etc., contain problems that are more difficult and competitive. In addition to the preceding function-level benchmarks, researchers have also explored LLMs with programming problems that are more complex but also more pragmatic.
ClassEval~\cite{ClassEval} is constructed to evaluate class-level code generation. CoderEval~\cite{Codereval} is constructed for the evaluation of non-standalone functions. RepoBench~\cite{RepoBench} and RepoEval~\cite{RepoEval} consider the evaluation of repository-level code auto-completion.

\textbf{Benchmarks of different applications.} There are some benchmarks concerning code generation for specific applications. 
Methods2test~\cite{Methods2Test} and Test4J~\cite{Test4J} care about the capability of generating test cases. 
SWE-Bench~\cite{SWE-bench} and HumanEval-Java~\cite{HumanEval-Java} evaluate LLMs on generating patches for existing programs.
AVATAR~\cite{AVATAR} and XLCoST~\cite{XLCoST} are constructed for code translation, facilitating the evaluation of cross-lingual code intelligence.

Unlike most existing studies using natural-language descriptions to present the target functionalities, our evaluation presents the target functionalities through only (iteratively supplemented) I/O examples. 
Similarly, Li et al.~\cite{PBELLM} evaluate the capabilities of large models on three domain-specific code generation tasks, but the examples that they provide to the LLMs are given all at once, still suffering from incomplete descriptions.
As the first work formalizing programming functionalities into iteratively supplementary I/O examples, our evaluation framework and benchmark can also be used to evaluate the capability to implement multi-turn-provided requirements.

%% file: Threats.tex
\section{Threats to Validity}

\noindent \textbf{Construct Validity.} 
The term \emph{example-based code generation} may also be associated with \emph{programming by demonstration}~\cite{demonstration}, where programmers (usually the end users) demonstrate operations on example data, and the computer records and generalizes these operations with programs.
In this paper, we restrict the definition of \emph{examples} to the input-output pairs of the target functionality. 
The evaluation takes pass@k as the primary metric, which mainly concerns correctness but does not fully reflect code quality, such as readability and maintainability. To enable further investigation in the community, we open-source all the generated code collected in our experiments.

\noindent \textbf{Internal / External Validity.}
(1) Dataset Bias: the distribution of (the types of) target functionalities is biased and may affect the final evaluation results, although the considered functionalities are all derived from widely used benchmarks for code generation. To mitigate this threat, in Section~\ref{sec: results} (RQ3), we separately compare different LLMs on each type of functionality.
(2) Model Configuration: different hyperparameters across different models may also result in variations in performance. In our evaluation, we adopt the default parameters for all the evaluated models and explicitly point out all other parameters used in the experiments.
(3) Tool Reliability: in our evaluation framework, due to the discrepancies in the C\# versions supported by tools, in very few cases, correct code can be determined as failures after compilation. To mitigate this threat, we manually check and correct the compilation results as thoroughly as possible.
(4) Generalizability: the evaluation considers only code written in C\# language, and the conclusions may not be applicable to other programming languages, especially those that are less widely used or are domain-specific. As one of the most commonly used programming languages in natural distribution, C\# is also widely considered in the evaluation for code generation~\cite{lozhkov2024starcoder2stackv2}. Therefore, we believe that the evaluation against C\# code is representative. Additionally, with acceptable engineering efforts and tool support, the evaluation framework in this paper can also be extended to support other programming languages.

\noindent \textbf{Conclusion Validity.}
The interpretation of results may be affected by subjective judgment, especially for those requiring manual inspections. To mitigate this threat, we involve multiple researchers to conduct the result analysis and cross-validation.

%% file: Conclusion.tex
\section{Conclusion}
In this paper, we have presented a comprehensive study of six large language models (LLMs) on example-based code generation.
We have found that GPT-4o-mini and DeepSeek-Coder perform the best, both in generating code that satisfies given input-output examples and in inferring the target functionality.
However, the LLMs still struggle when the target functionality is defined solely through input-output examples, because of the difficulties in both understanding the requirements and in effectively using iterative feedback.
We also discussed the impact of the type of target functionalities, the selection of input-output examples, and the introduction of natural-language descriptions, as an exploration of potential improvements.
Through the comprehensive assessment and analysis, this study reveals the limitations of LLMs on example-based code generation, calls for more support of diverse forms of requirement descriptions, and emphasizes the importance of early prompts in the code generation tasks described through iterative conversations.

\section{Data Availability}
We open-source our data and evaluation results on our project website~\cite{website}.
